# SEQUENTIAL RECOMMENDATION MODEL FOR NEXT PURCHASE PREDICTION

Xin Chen, Alex Reibman, Sanjay Arora

Ernst & Young LLP. US.

## ABSTRACT

*Timeliness and contextual accuracy of recommendations are increasingly important when delivering contemporary digital marketing experiences. Conventional recommender systems (RS) suggest relevant but time-invariant items to users by accounting for their past purchases. These recommendations only map to customers' general preferences rather than a customer's specific needs immediately preceding a purchase. In contrast, RSs that consider the order of transactions, purchases, or experiences to measure evolving preferences can offer more salient and effective recommendations to customers: Sequential RSs not only benefit from a better behavioral understanding of a user's current needs but also provide better predictive power. In this paper, we demonstrate and rank the effectiveness of a sequential recommendation system by utilizing a production dataset of over 2.7 million credit card transactions for 46K cardholders. The methodfirst employs an autoencoder on raw transaction data and submits observed transaction encodings to a GRU-based sequential model. The sequential model produces a MAP@1 metric of 47% on the out-of-sample test set, in line with existing research. We also discuss implications for embedding real-timepredictions using the sequential RS into Nexus, a scalable, low-latency, event-based digital experience architecture.*

## 1. INTRODUCTION

Recommender systems (RS) suggest relevant items to users by accounting for preferences and past purchases. A RS can narrow down purchase options by marketing attractive items andthereby enhance a user's experience and boost sales. In terms of business value, RSs have been shown to increase click through rates (CTRs) by up to 30%, lift adoption and conversion rates by 2-5%, and grow sales revenue [1].

Recommendations in e-commerce emerged in popularity in the early 1990s [2], and they are increasingly being contextualized to a user's recent behaviors, such as purchase or browsing history. The need for a certain product or service, for example an out-of-town meal, may result from recent life events (e.g., a career change) that may in turn be associated with similar transactions (e.g., a hotel stay and airfare). But after a certain period, evolving changes in lifestyle may no longer be influenced by past purchasing behaviors. The goal of recommenders in this context is to anticipate time-sensitive, emergent user needs and surface products, offers, or merchants to address these needs.

Viewing data as a sequence of events – and as a longitudinal set of behavioral indicators – has changed the way researchers and practitioners view recommendation systems. Advances in deep learning have also changed the way recommender systems are built and measured across different domains [3]. For example, music and video recommenders use sequences of audio and images to help isolate defining features of the media and associate those features with user preferences. Some natural language processing methods, on the other hand, read character, word, or phrase tokens and output translations or new chunks of text as output.

One challenge with sequential recommenders in digital marketing and e-commerce is the ability to bring together all necessary data (user, item, and contextual) into an architecture capable of serving insights to users in just the right moment. In other words, it's not enough to just train a performant model: That model must also scale with large datasets, deliver recommendations with low-latency, and manage long-term data dependencies, feedback loops, testing strategies, and concerns about fairness [4, 5].

This paper documents the development of an offers recommendation system using sequential models and highlights the componentry necessary to provide the model's insights to end users. Accordingly, we offer three main contributions:

1. A sequence and context aware deep learning model capable of delivering high-quality user recommendations.
2. An evaluation framework that positions our models' performance near the top of other similar models.
3. A high-level architecture for delivering predictions to end users through EY's Nexus product platform.

### 1.1. Use Case

User-item interactions are often sequentially dependent. For example, the purchase history in Figure 1 shows a user booked a flight, a taxi, and a hotel successively, and will likely visit restaurants in town. In this case, a user's purchase behaviors depend on his or her last few purchases. In addition, individual preferences change over time and item attributes (e.g., price, style, brand) may vary in importance at different times. For instance, an individual who used to eat a lot of fast food may have now decided to live a healthier life. Their diet changed, but perhaps they also started to go to the gym and make additional purchases on fitness equipment and apparel. These dynamic changes are of great significance to future recommendations presented to the user. Indeed, presenting fast food and stationary activities may no longer align to new life goals. Lastly, recommendations are only relevant if they are attainable to the user, so the RS must also include a re-ranking step to only present merchants within a reasonable distance from the user, or within a reasonable price point.

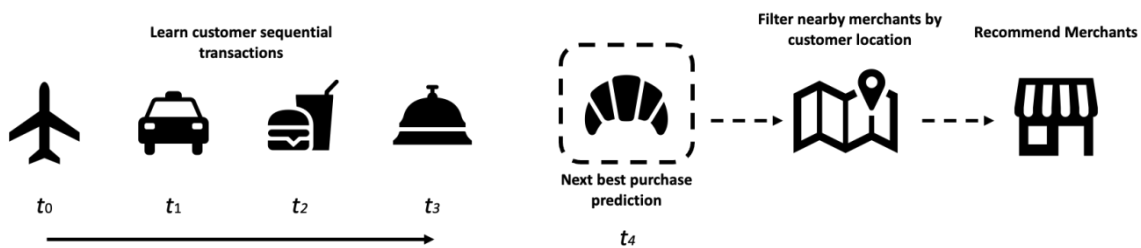

Figure 1. Identifying qualifying merchants based on customer reservation price preferences, sequential learning recommendations, and location filtering.

It is challenging for traditional recommendation systems to capture these sequential dependencies and dynamic preference changes. Sequential recommendation systems (SRSs), in contrast, take a sequence of user behaviours as input and capture both long and short-term interests to recommend the next most likely purchased items more accurately.

### 1.2. Related Work and Recent Progress

Conventional recommendation systems, including content-based and collaborative filtering RSs, make predictions based on the user's past purchase behavior and/or decisions made by other users with similar preferences [2, 6]. Content-based filtering typically relies on item-specific qualities, e.g., the properties of a merchant, product, song, or webpage, whereas collaborative filtering approaches recommend items that similar users have liked, browsed, or purchased previously. Hybrid approaches combine aspects of recommendation techniques, for example aiming to minimize disadvantages of collaborative filtering and content recommenders by considering both user similarity and item information. Regardless of type, a conventional RS assumes all historical user-item interactions are equally important; therefore, they can only capture the users' static and general preferences. Additionally, item information typically remains static, meaning that the relationship between one item and another does not dynamically change over time. However, as noted previously, preferences, needs, and life situations often change the saliency of certain recommendations, and therefore sequential RSs (SRSs) are gaining prominence in both academic and practitioner circles [3, 7, 8].

Figure 2 summarizes reported algorithms for SRS. Early works using Markov chain (MC) models for next action prediction consider the past behavior sequences [9]. The MC method highly depends on past occurrences and calculates the probability of the next state only by depending on the present state with first order MC [10]. Though high-order MC models take additional previous behaviors into consideration [10], it is still a challenge for MC to deal with long-term sequences, new items and contextual information. In particular, higher-order MC models suffer from two main limitations [11]: Firstly, although earlier actions can independently influence later actions, they cannot jointly do so. For example, purchasing a train ticket and hotel room increases the probability of an out-of-town meal more than either event alone. A second, related limitation of Markov chain-based models is the inability to "skip over" actions that are unrelated to the prediction of the next most likely action. Sequential deep learning-based models can address both of these limitations. That said, most recommendation models, including sequential ones, still are limited by common data availability problems (e.g., data sparsity and cold start) [3, 6, 11].

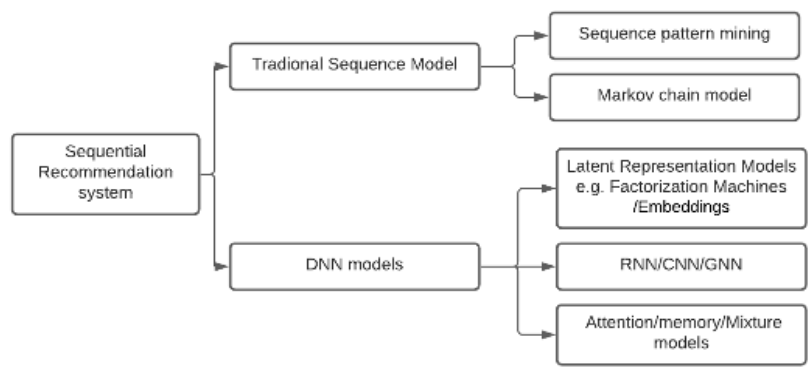

Figure 2. SRS approaches from the technical perspective.

More contemporary approaches apply deep-learning based sequential recommendation techniques. This latent representation technique first learns representational embeddings for each user and item and then utilizes matrix factorization to predict the probability of user-item interactions for the recommendation [12]. Sequential neural networks, including RNN, GRU, and LSTM, have been developed to capture the long-term dependencies in a sequence for recommendations in a single domain [13]. For instance, convolutional neural networks (CNNs), though commonly used in the computer vision field, have been applied in sequential movie recommendations [14]. The input is a sequence of embeddings where each embedding represents a watched movie in the past. The CNN-based model treats the sequence of embeddings as an image and learns sequential features by convolving over the "image". Other more complex deep learning structures are used in SRSs, including Graph Neural Networks (GNNs), attention-based models, and memory networks. However, most of these sequential models are for single domain recommendations [3, 12].

Several deep learning-based SRSs, including GRU4Rec [15], C-GRU-based [16], P-GRU-based [17], CNN-based (Caser) models [11], have been reported that outperform conventional recommender systems (e.g., POP, KNN, BPR-MF) by 20-30% on multiple public datasets (e.g., RSC15, RSC19, MovieLens) [3]. More complex structures, i.e., attention incorporated RNN model, further improves SRS performance by 25% [3]. In addition, Transformer architecture has also been applied to predict user's next action by capturing user's preferences from user's past action sequences [18, 19].

The rest of the paper is organized as follows: section 2 will describe the proposed sequential recommendation model in detail, including the overall architecture diagram and components of the proposed model, and how the models are trained and used to predict the next purchase. The experiment process, transaction dataset used for training and testing, model evaluation, and experiment results will be presented and discussed in section 3. Experiment results will show how the mode benefit from the auto-encoder transaction encoding as well as additional context information. Section 4 will give a high-level overview of architecture for delivering predictions to end users through EY's Nexus product platform. Finally, we conclude our work in section 5.

## 2. PROPOSED METHODOLOGY

We propose a sequential recommendation model SEQuential recommendation model for Next Best Transaction (SEQNBT) for predicting the industry code and transaction amount for a user's next most likely transaction. Figure 3 depicts SEQNBT's structure. It consists of an auto-encoder, GRU, and transaction decoder. The auto-encoder model is a self-training model used for generating transactional encodings, which are used as an input to the GRU model to capture sequential dependencies and users' long-term preferences. Assume a set of users $U$ ($u_1, u_2, u_3, ... u_M$), and a finite set of industry category code (SIC) $I$ ($i_1, i_2, i_3, ... i_N$): each user has made several transactions within different industry categories from set $I$ in the past. We use $S$ to denote the transaction sequence a user made in chronological order $S$ ($s_1, s_2, s_3, ... s_t$), where $s_t$ indicates the transaction user made at a relative time step $t$ [11]. SEQNBT takes the user's past $L$ transaction sequence as input and predicts the SIC code and transaction amount of his/her next most likely transaction. Note that the raw recommendations in our model are industry classes (SICs) along with transaction amounts. SICs get mapped to specific nearby and/or affordable merchants to provide business value to end-users.

Our models are trained on a real-world credit card transaction dataset. Each raw transaction record contains the user's hashed card holder number, transaction date, transaction sequence number, net billed amount, supplier number, merchant category code, industry category code, and industry category description. Detailed information about the transaction dataset will be introduced in Section 3.1.

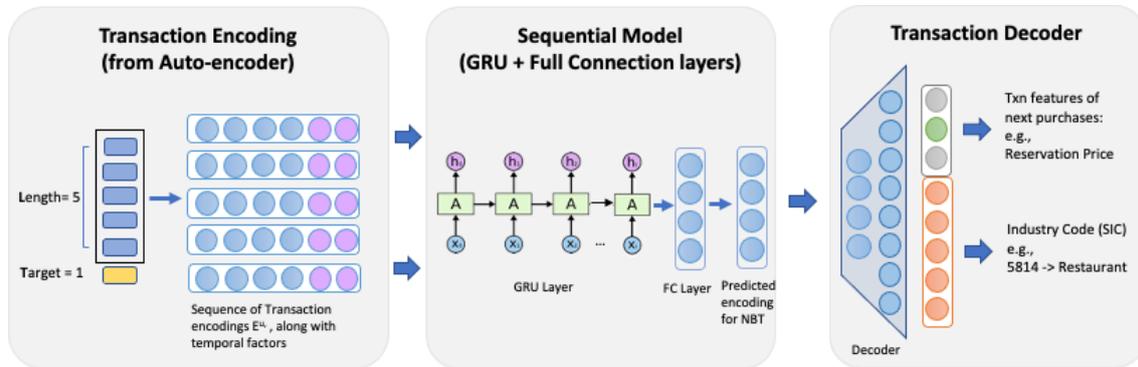

Figure 3. Architecture diagram of SEQNBT model.

## 2.1. Auto-Encoder

Most existing work utilizes an embeddings matrix [20, 21] or 1-of-N encoding [15] to represent items (industry code of transactions in our case) that users interacted with. However, extra item information, such as text descriptions and item images, can also be exploited to improve recommendation performance [3], especially when users' profiles are not available [17]. To better represent users' credit card transactions, we make use of additional information, including textual descriptions of each industry and transformed features of each transaction; from there, we apply a stacked auto-encoder to extract efficient representations of each transaction.

A stacked auto-encoder is an unsupervised neural network model capable of learning a dense representations of input data [22]. Figure 4 depicts the architecture diagram of the stacked auto-encoder model. It always consists of two parts: An encoder that maps inputs to a latent representation (encoding), and a decoder that reconstructs the latent representation back to an output in the input space [22]. The encoder part and decoder are both symmetrical to the central encoding layer (bottleneck layer).

We performed extensive feature transformation and generated aggregated features on the user's overall spending behaviour (e.g., number of transactions, mean/median/total transaction amount), the user's spending behaviour (e.g., reservation price, the highest price a user is willing to pay for a product) within different merchant categories (e.g., restaurant, transportation, hotel), and on each merchant across all users (e.g., total number of transitions, mean/median/total transaction amount). We also used a pre-trained **B**idirectional **E**ncoder **R**epresentation from **T**ransformers (BERT) model to generate embeddings of SIC descriptions. The pre-trained BERT model, *bert-base-uncased*, is made available by Hugging face, an AI community provider of open-source NLP technologies. Hugging face offers thousands of pre-trained models to perform tasks, such as text classifications, question answering, translation, etc. [23, 24]. All aggregated transaction features along with transaction amount and SIC embedding are concatenated and fed into the encoder and reconstructed through the decoder. As with all stacked encoders, the difference between original input and reconstructed output are optimized during training.

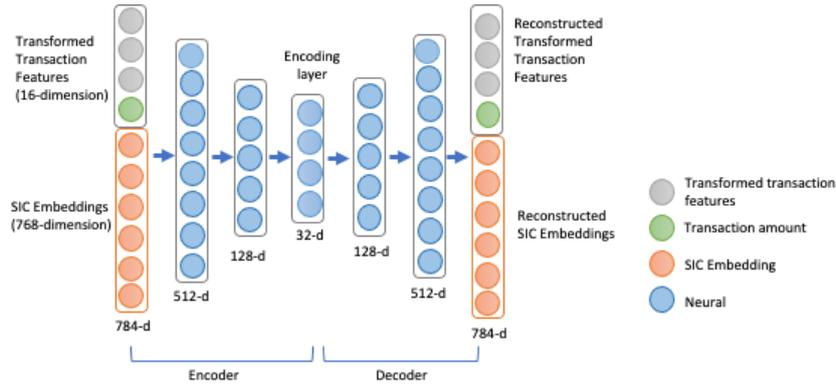

Figure 4. Architecture diagram of auto-encoder for transaction encoding

After obtaining the trained auto-encoder model, each transaction that user made previously is encoded as a $d$- dimension vector ($d = 32$ in our case) by the encoder model. Therefore, to predict user's next most likely transaction, a user's previous L transactions at time point $t$ are represented as an encoding Matrix $E_t^u$ ($L \times d$) as equation (1), acting as the primary input to the SEQNBT model. $e_t^u$ is 32 - dimension encoding for a transaction made by the user $u$ at time step $t$. The user's next transaction encoding ($e_{t+1}^u$) will be used as target value. The decoder is used to decode the predicted transaction and retrieve the original industry code and transaction amount for the next most likely transaction a user is likely to make. The detailed decoding process will be illustrated in Section 2.4.

$$E_t^u = \begin{bmatrix} e_{t-L+1}^u \\ \ldots \\ \ldots \\ e_{t-1}^u \\ e_t^u \end{bmatrix}$$

→ [0.9 0.6 0.1 0.3] Encoding of user' previous L transaction made at time t-L+1
...
→ [0.4 0.1 0.7 0.1] Encoding of user' previous 2 transaction made at time t-1
→ [0.2 0.3 0.2 0.3] Encoding of user' previous 1 transaction made at time t

(1)

## 2.2. Gated Recurrent Unit (GRU) layer

Recurrent Neural Networks (RNN) are a class of neural networks that model sequence data and make predictions by considering both the current input as well as what it has learned from the inputs it previously received. RNNs usually have short-term memory and are in capable of capturing long-term dependencies because of gradient vanishing or exploding, (i.e., the gradient is decreasing or increasing exponentially with respect to the number of neural layers and never converges to the optimum). GRU solves these issues by introducing two vectors: an update gate and reset gate, which can be trained to determine (a) what information should be kept without vanishing through time and (b) what information is irrelevant to the prediction that can be removed. GRU has been widely applied in SRSs [15, 17, 25]. Our proposed SEQNBT model is based on multiple GRU layers to capture dependencies and user preferences from past transaction sequences and to output a next most likely transaction.

The input of these GRU-based models usually consists of item embeddings, and / or user's embeddings, and / or item category embeddings [16], but temporal features are often neglected. Temporal features have a significant impact on the user's purchase behavior. Therefore, in addition to encodings generated from the auto-encoder to represent user's transactions, we also consider time factors in our proposed SEQNBT for next most likely transaction prediction. The temporal features (marked in lavender circle in "Transaction Encoding" box in Figure 3) include day of week, day of month, whether the transaction was on a weekend, and the time delta from previous transactions in days; these features are generated and concatenated with the aforementioned encoded sequence (marked in blue circle in "Transaction Encoding" box in figure 3). More formally, the input sequence of SEQNBT model for a user $u$ at time step $t$ is denoted as $ES_t^u$, in which $e_t^u$ is the transaction encoding of the user $u$ at time step $t$, and $temp_t^u$ indicates temporal features of the transaction the user made at time step $t$. The encoded transaction plus the temporal features is then fed into the SEQNBT model to obtain a vector representing the next most likely transaction.

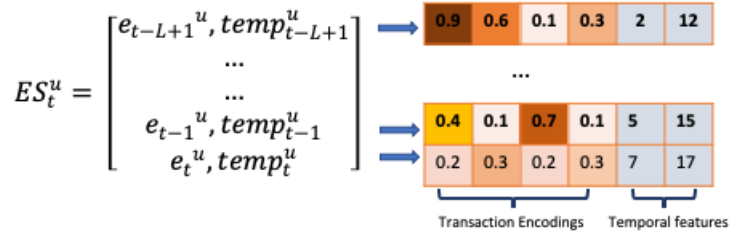

$$ES_t^u = \begin{bmatrix} e_{t-L+1}^u, temp_{t-L+1}^u \\ \ldots \\ \ldots \\ e_{t-1}^u, temp_{t-1}^u \\ e_t^u, temp_t^u \end{bmatrix} \quad (2)$$

## 2.3. Model Training

### 2.3.1. Auto-encoder training

The auto-encoder model is trained separately from SEQNBT model. We used a stacked auto-encoder with symmetrical architecture regarding the central hidden layer. To train the auto-encoder model, the pre-transformed aggregated transaction features (marked in gray circles in Figure 4) concatenated with transaction amount and BERT embedding of SIC description of each transaction are used as inputs, and the encoder deterministically maps the input vector (784-dimension) to a hidden representation by:

$$h_i = a(Wx_i + b) \quad (3)$$

where $W$ and $b$ are weights and bias of encoder, and $a$ is the activation function Scaled Exponential Linear Unit (SELU) denoted as:

$$a = \lambda \begin{cases} 1 & if\ x > 0 \\ \alpha e^x & if\ x \leq 0 \end{cases} \quad (4)$$

The decoder reconstructs the resulting hidden representation $h_i$ back to 784-dimension vector $y_i$ in input space as:

$$y_i = a(W'h_i + b') \quad (5)$$

where $W'$ and $b'$ are weights and bias of decoder. We used a custom loss function that calculates the sum of mean square error of the original and reconstructed transaction features ( $\|x1_i - y1_i\|^2$ ), as well as the mean square error of the original and reconstructed SIC embeddings ($\|x2_i - y2_i\|^2$):

$$L(x, y) = \frac{1}{n}\sum_1^n(\|x1_i - y1_i\|^2 + \|x2_i - y2_i\|^2 \quad (6\text{-}1)$$
$$x = [x1, x2],\ y = [y1, y2] \quad (6\text{-}2)$$

where $n$ is the total number of training samples. The loss function is minimized to find the optimal parameters $\theta = \{W, W', b, b'\}$ during the training process. The hyperparameters, including the number of encoder layers, number of units in each encoding layer, activation functions in each layer, batch size, and learning rate, are tuned by sequential search of hyperparameter space via Hyperopt on a validation dataset. Hyperopt is an open-source python library for optimizing hyperparameters of machine (deep) learning models [26]. We utilized the Adam optimizer for efficient computation. Once the auto-encoder model is trained, the encoder model is used to generate transaction encodings for every single transaction in our dataset. The transaction encodings are stored in a specialized database called a Feature Store. Encodings are saved in an Offline Store for lookups while training the SEQNBT model, and they are also saved in an Online Store used for fast lookups data when running real-time scoring in a production environment.

Table 1: Optimized hyperparameters of auto-encoder

| Model | Batch Size | # Of encode layers | # Of units in encoding layers | Learning rate | Epoch |
|---|---|---|---|---|---|
| Auto-encoder | 128 | 3 | 512 (layer 1)<br>128 (layer 2)<br>32 (layer 3) | 5e-5 | 100 |

### 2.3.2. SEQNBT Model Training

For each user, the transaction encodings are ordered chronologically by transaction date and transaction sequence number. We extract training observations using a sliding window of size $L + 1$ over the user transaction encodings, to have $L$ successive transaction encodings as input and the +1 encoding as the target to be predicted. The sliding window moves 2 steps each time to avoid large overlaps on transaction sequences between observations of the same user.

To train the SEQNBT model, the encoding sequence of size $L$ along with temporal features are fed into GRU layers and a fully connected layer to predict a 32-dimension "encoding" of the next transaction. The mean squared deviation between the predicted "encoding" and the target transaction encoding is to be minimized. We adopt an Adam optimizer with batch size of 200. In this paper, we follow the best practices in other research that use similar learning rate decay and dropout methods [15, 22]. The learning rate is scheduled to decay exponentially with initial learning rate of 0.001 and decay rate of 0.9 every $10^5$ steps. A dropout layer with dropout ratio 0.25 is applied following each GRU layer except for the last GRU layer, to avoid over-fitting.

We optimized the hyperparameters, including number of GRU layers, number of RNN units in GRU layer, epochs, decay rate and batch size, by running 50 experiments at sequential selected points of the parameters space using hyperopt. The best hyperparameters of SEQNBT model are summarized in the Table 2.

Table 2: Optimized hyperparameters of SEQNBT

| Model | Batch size | # of GRU layers | # of units in GRU layer | Dropout rate | Learning rate | Epoch |
|---|---|---|---|---|---|---|
| SEQNBT | 200 | 3 | 64 (same for all GRU layers) | 0.25 | 0.001 | 60 |

### 2.3.3. Transaction Decoder

Once the SEQNBT model was trained, we evaluated SEQNBT on an out-of-sample testing dataset to predict encodings for each user's next predicted transaction. A predicted transaction encoding (32-dimension) is then decoded by the decoder model into two parts: (1) the concatenated transaction features vector (16-dimension), and (2) the SIC embedding vector (768-dimension). We calculated cosine similarity between predicted SIC embedding and all SIC embeddings in the feature store table storing BERT representations for each SIC; the result of these comparisons is an SIC number ranked by cosine similarity score. Predicted transaction amount is among the decoded transaction features (shown as a green circle in Figure 3).

## 3. EXPERIMENTS

We compared our proposed model with a set of state-of-the-art models, including GRU4Rec [15], a GRU-based model to recommend next item in a session, and AttRec [27], a self-attention based model for next item recommendation, on our real-world banking transaction dataset. These two models, GRU4Rec and AttRec, do not take into account item context but rather employ user and item embeddings. Source code for GRU4Rec and AttRec are available online [28]. We also studied the effect of sequence length $L$ on model performance.

Table 3. Banking transaction dataset

| Columns | Description | Data sample | Unique numbers |
|---|---|---|---|
| Hashed_Card_Number | Hashed cardholder number | e6e2fc…74 | 46006 |
| Transaction date | Date that transaction made | 2018-09-30 | 148 |
| Trans_Seq_No | Transaction sequence number | 481816 | 1294652 |
| Net_Billed_Amount | Transaction amount | $36 | - |
| Supplier_Number | Supplier number | 5430413435 | 186377 |
| MCC_No | Merchant Category | 5812 | 263 |
| SIC_No | Industry Code | 5814 | 290 |
| SIC_String | Description of SIC number | Fast Food Restaurants | 290 |

### 3.1. Dataset Description

We conducted our experiments on business travel credit card transaction data, containing 2.7M purchase transactions for 46K cardholders at 186K merchants within 251 industry categories from September 30, 2018, to March 31, 2019. Each raw transaction contains hashed cardholder number, transaction date, transaction sequence number, transaction amount, supplier number, merchant category, industry category number, and industry category description. The raw transaction data is shown in Table 3.

We pre-processed the raw transaction dataset by removing users who made fewer than 10 transactions at fewer than 5 distinct merchant categories to avoid the cold-start issue [27]. Descriptive statistics before and after processing are summarized in Table 4. The density of a dataset is defined as the ratio of total number of interactions and total number of potential interactions between users and items. The density of our transaction dataset is about 34.42% before pre-processing, which is much denser than other commonly used public datasets, whose density is often less than 7% [27, 29]. We observed our model outperforms the two baseline models on our dense transaction dataset; we discuss this further in Section 3.1.2.

$$Density = \frac{total\ number\ of\ interactions}{number\ of\ users * number\ of\ items}$$

Table 4: Statistics of the datasets used in experiments.

| Datasets | Before processing | | | | After processing | | | |
|---|---|---|---|---|---|---|---|---|
| | # of users | SIC code | # of transactions | Density | # of users | SIC code | # of transactions | Density |
| Banking transaction dataset | 113011 | 272 | 9965631 | 32.42% | 40006 | 252 | 2696143 | 26.74% |

## 3.2. Evaluation Metrics

We adopt the commonly used evaluation measures of mean average precision (MAP) and recall at varied cut off values (K = 1, 5, 10) to evaluate the performance of the SEQNBT model. AP (averaged precision) is the sum of precision@k for different value of k ranging from 1 to $K$ and divided by total number of target items. It is defined as:

$$\text{AP@K} = \frac{1}{T}\sum_{k=1}^{K} P(k) \cdot rel(k) \tag{7}$$

where $T$ is total number of target items ($T = 1$ throughout the paper). $P(k)$ is the precision of the k$^{th}$ recommendation, and $rel(k)$ indicates whether the k$^{th}$ recommendation was relevant ($rel(k) = 1$) or not ($rel(k) = 0$). MAP@K is averaged AP@K across all users. Figure 5 demonstrates an example of calculating MAP@K (K = 5). Higher values approaching 1 for MAP@K are better, whereas 0 signifies no matched industry code. Note that mean reciprocal rank (MRR) at K, an another commonly used evaluation matrix, leads to the same results as MAP@K in terms of predicting the (one) item; therefore, we do not include MRR in our work.

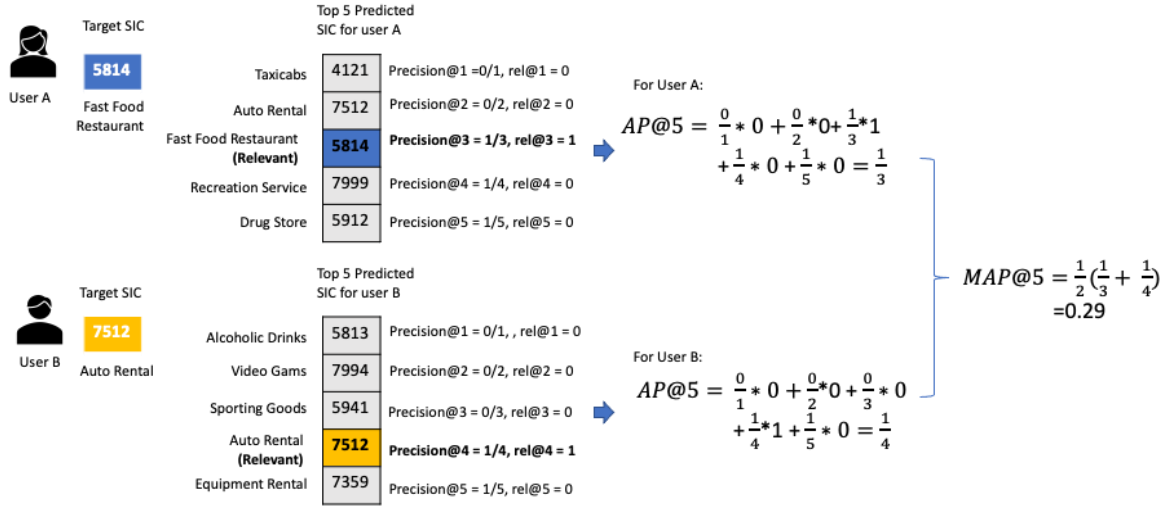

Figure 5. Example of how MAP@K (K = 5) is calculated. Note a hit always has one in the numerator. Values can range between 0 and 1.

Recall@K evaluates the proportion of desired (target) items that are among the top K predictionsin all test cases, without considering the rank of items in the prediction list. It is defined as the ratio of the total number of relevant items among the top K predictions to the total number of target items. Recall@K for each user is either 0 or 1 in the case of predicting the next (one) item:

$$\text{Recall@K} = \frac{|Target \cap Pred[1:K]|}{|T|} \tag{8}$$

Averaged Recall@K across all users is calculated to evaluate model performance.

## 3.3. Model Analysis and Discussion

We evaluated the proposed SEQNBT model against two popular baseline models, GRU4Rec and AttRec, on our banking transaction dataset. The two baseline models are first trained to reproduce the performance reported in these two papers [11, 27] on the MovieLens-100K dataset, i.e., before being tested our banking transaction dataset. Note we only train the GRU4Rec and AttRec on the MovieLens dataset to extrapolate hyperparameters (e.g., number of layers, learning rate) for use on our credit card dataset. The hyperparameters used in the two baseline models are summarized in Table 5. We also studied the effect of sequence length on performance of SEQNBT model by varying sequence lengths from 3, 5, and 7 to 10.

Table 5. Hyperparameters used in the baseline models.

| Model | Batchsize | Embedding size | Negative samples | Learning rate | Loss function | L2lambda | Epochs |
|---|---|---|---|---|---|---|---|
| GRU4Rec | 256 | 50 | 10 | 0.01 | 'top1' | 1e-6 | 30 |
| AttRec | 100 | 100 | 1 | 0.001 | margin-based hinge loss | 1e-6 | 35 |

### 3.3.1. Impact of Sequence Length on Performance

The impact of sequence length on SEQNBT model performance is presented in Figure 6. The performance of SEQNBT increases as sequence length increases from 3 to 5, then drops as sequence length gets longer (e.g., to 7 or 10). This may occur because, if the length of input sequence is too short (e.g., $L = 3$), then there is not enough data to track a user's preferences, especially when the user made three transactions in the same industry category (e.g., a user made three successive transactions in the same restaurant category). The performance reduction when the sequence length is long ($L > 5$) may result from a smaller number of training samples, a finding which is aligned with previous work [3].

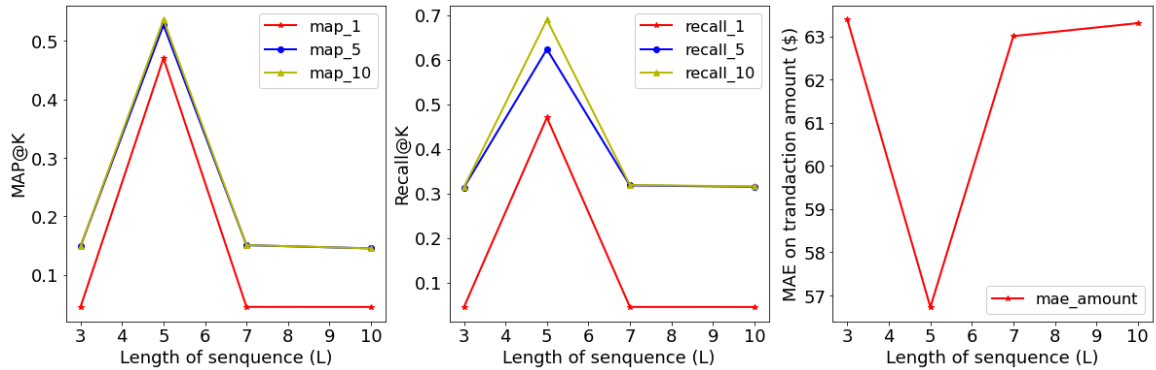

Figure 6. Impact of sequence length L on MAP@K, Recall@K (L = 3, 5, 7, 10; K = 1, 5, 10)

Sequence length exhibits a similar effect on prediction of transaction amount. The lowest mean absolute error (MAE) on predicted transaction amount we obtained so far is $56.73 on the testing dataset with length of input sequence of 5. We observed our model performs slightly better on predicting transaction amount of the next most likely transaction for low-spend industry categories (e.g., retailing stores, taxis, restaurants) than for high-spend industry categories (e.g., airlines, hotels).

The MAE of predicted transaction amount is positively correlated (0.72) with averaged transaction amount across all SIC categories, as shown in Figure 7. This may result from the imbalanced nature of our transaction dataset, where around 59% of transactions are within the low-spend industry category of "restaurant" (31%) and "taxicab" (28%), while transactions within high-spend categories are less frequent. Future work may improve upon the accuracy of predicted transaction amount by: (1) collecting (or performing data augmentation on) more transactions within high-spend categories; (2) optimizing the SEQNBT model specifically on transaction amount.

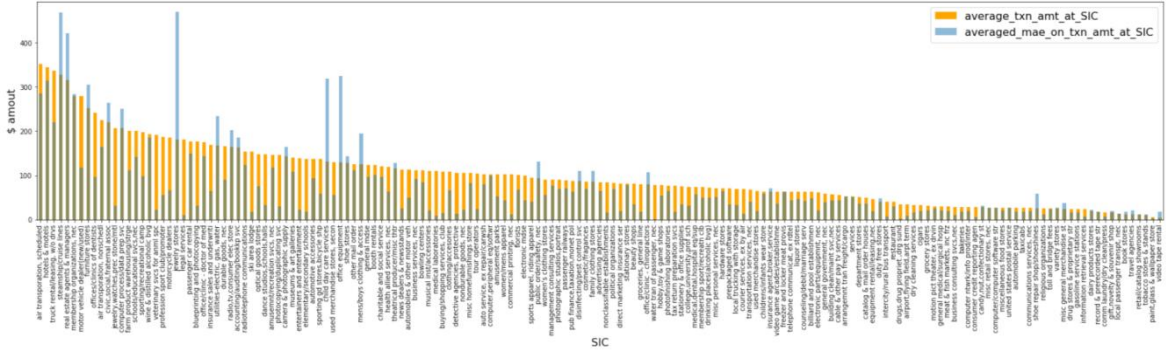

Figure 7. Comparison between averaged MAE of predicted transaction amount and averaged transaction amount across industry categories (SIC). The blue bar represents the averaged MAE on predicted transaction amount at each SIC category, and the orange bar represents the average transaction amount at each SIC category.

### 3.3.2. Performance Comparison with Baseline Models

Table 6 summarizes the experimental results of our SEQNBT model and two baseline models on our banking transaction dataset. The 2 baseline models are pretrained on MovieLens-100K datasets to ensure the model can reproduce the results reported in previous work [15, 27]. The length of input sequence and target is five and one, respectively for all the tests. Our model outperforms the other two baseline models in both prediction accuracy and ranking quality. We attribute the superior performance of the SEQNBT model vis-à-vis AttRec and GRU4Rec due to considering extra item context, including SIC embeddings, as well as temporal information. Previous research found that item context may yield better results than just user and item embeddings [3, 17], which are central to the AttRec and GRU4Rec approaches.

Table 6: Comparison between SEQNBT and baseline models on banking transaction datasets

| Model | MAP | | | Recall | | |
|---|---|---|---|---|---|---|
| | @1 | @5 | @10 | @1 | @5 | @10 |
| SEQNBT | 0.4704 | 0.5274 | 0.5363 | 0.4704 | 0.6239 | 0.6901 |
| AttRec | 0.0273 | 0.0408 | 0.0434 | 0.0273 | 0.0657 | 0.0851 |
| GRU4Rec | 0 | 0.0131 | 0.018 | 0 | 0.0346 | 0.0724 |

[a] Sequence length is 5, and target length is 1.

## 4. ML ARCHITECTURE

ML code is only a small fraction of a real-world ML system: The rest of an ML system relates to configuration, data sourcing, feature extraction, data verification, server management, process management, serving management, and model monitoring [4]. ML systems, as a consequence, not only share design, build, and maintenance challenges with traditional software systems but also introduce new challenges. Servan and Visser [30] highlight a broad array of ML issues, including (1) insufficient functional requirements; (2) the need to abide by some type of regulation – without knowing how to explicitly satisfy regulators; (3) a data sourcing 'jungle' of pipelines for model training and scoring; (4) comingling of code dependencies across logically distinct phases of the AI/ML lifecycle; (5) high coupling between components of the ML architecture, with cascading effects of errors; (6) testing; (7) model monitoring and the need to retrain models to protect against drift; (8) balancing latency, throughput, and fault-tolerance; and (9) tapping multiple and often hard-to-source skill sets [31].

The use case presented above for next-best transaction prediction requires complex data processing for the SIC BERT embeddings, transaction encoder, sequential model, and transaction decoder. In total we maintain five separate models, aggregate data at the user and merchant levels, employ raw transaction features (e.g., transaction amount), and store reference data and intermediate model outputs in a *feature store*. In addition, our model outputs are persisted to an *insights store* to track consumer-ready predictions that can then be used as labeled data for model retraining and continuous training purposes.

Our models and ML architecture assets are part of EY broader product platform, Nexus. Nexus help enables the world's leading financial services firms to re-imagine, create, and run better ways of providing growth in the digital economy. The platform provides a versatile, cloud-ready framework, with ready-to-use building blocks and real-time intelligence. Nexus provides customers with personalized experiences, at the right time, in the right channel to demonstrate situational awareness and orchestrate both front-end journeys and back-end business processes. The platform provides these capabilities by capturing events, processing triggers in real-time, generating insights and nudge actions, harnessing customer responses, and in this use case context, implementing marketing campaigns. Nexus builds on the growing popularity of event-driven architectures, which in terms of industry adoption according to one recent survey, are followed by lambda, microservices/SOA, layered, and workflow approaches [30].

Figure 8 shows Nexus' logical architecture. The underlying component enabling the Nexus platform is its *event engine*, which translates incoming signals into directed acyclic graph (DAG) *event maps*. Depending on the specific context, these events are used to kick-off ETL or feature generation pipelines, scoring procedures (e.g., the SEQNBT framework), customer nudges, etc. The *insight engine* facilitates model execution and supports feature store lookup, communication with downstream services, and sidecar patterns, such as logging and audit. Lastly, the *customer experience management* (CEM) module is responsible for deciding which messages to send, i.e., arbitrating amongst multiple, competing use-case driven communications to best support enterprise business objectives and to not overwhelm the end-user.

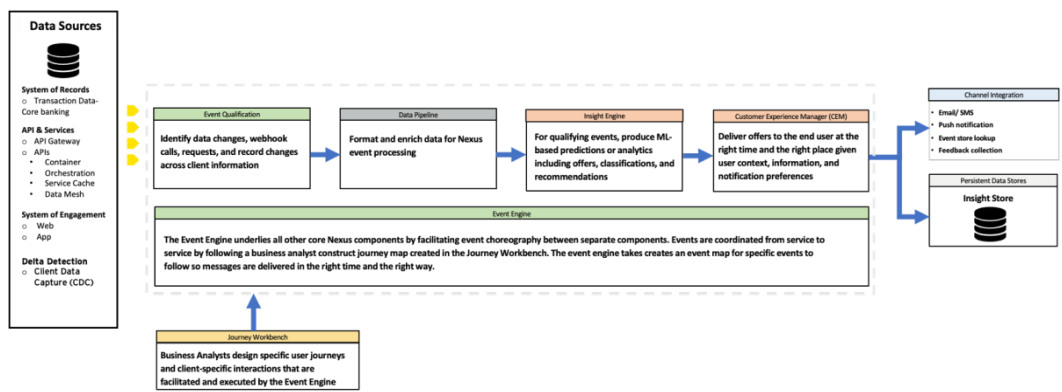

Figure 8. Nexus Platform Architecture

We use Nexus to embed the recommendations into offers-based journeys: That is, as soon as a customer enters a qualifying location, we can instantaneously update his or her available offers. While the offers themselves are identified by our SEQNBT and related models, specific delivery and notification preferences are handled by CEM. Nexus, then, is designed to facilitate the following: (1) real-time customer location tracking, (2) regularly scheduled user preference updates, and (3) real-time offer generation, delivery, and qualification. The ML architecture is largely supported by a message queue event handler that first identifies and qualifies customer locations. If the customer has entered a new location, then a set of supporting micro services requests a newly generated set of recommend industries from SEQNBT. SEQNBT is regularly retrained on a scheduled basis on a regularly refreshed set of credit card transactions. Lastly, once we have a list of likely industries for the next most likely transaction, and a list of qualifying merchants in those industries, we perform a re-ranking process whereby only merchants within a nearby vicinity of the customer are recommended. Figure 9 shows the detailed scoring pipeline, while Table 7 provides corresponding descriptions for each service in Figure 9.

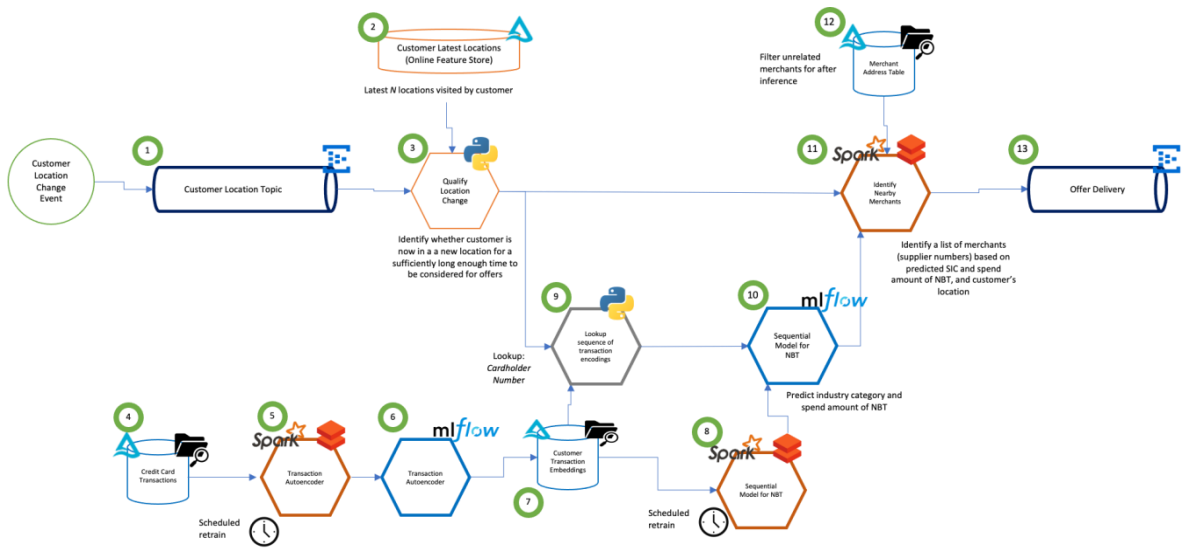

Figure 9. Nexus ML offers service architecture.

Modern ML architectures such as Nexus are emerging as the gold standard of contemporary insight service, which should be real-time, contextualized, and measurable in terms of efficacy and business impact. Nike recently published a survey of their recommendation system work; one model, ALSNN, had "won nearly all A/B tests with respect to key APIs, such as revenue-per-viewer and has generated millions of dollars in incremental revenue" (pp. 530) [32]. Nike combines multiple expert and stacked models to leverage several recommendation strategies, even addressing non-personalized approaches to solve the cold-start problem.

While Nike's recommendation model is tailored to its business model and core offerings, we anticipate that SEQNBT will work well across different product, digital, and financial data domains: Future work will necessarily entail testing the general SEQNBT framework with interaction and other recurring behavioural data to ascertain how well the framework generalizes. Additionally, we will continue to use the Nexus architecture to deploy, test, and monitor any future iterations of SEQNBT. For example, one planned revision at the time of writing is to incorporate self-attention into SEQNBT to calculate transaction sequence attention scores, which will help to identify more relevant items [13].

Table 7: Nexus ML offers service architecture – component descriptions.

| Id | Description |
|----|-------------|
| 1 | As customers change geolocation or emit events to the Nexus platform with geolocation data, the architecture will intelligently identify customers who should qualify for offers based on their current location. |
| 2 | For each customer, we store their latest location in a low-latency lookup specialized datastore called an *online feature store*. This is used to compare a use's current location against where they have been previously so we can deliver location-aware offers. |
| 3 | Qualify Location Change is a microservice to identify the customer's current location and whether they are currently qualified to receive an offer based on their location. |
| 4 | Datastore contains customer transactions at merchants. This data is fed into the autoencoder. |
| 5 | Transaction Autoencoder is used to re-construct transaction feature data. This microservice runs on a schedule and creates a serialized model file. |
| 6 | Deployed autoencoder trained in the previous job is used to do prediction at real-time. |
| 7 | Transaction embeddings from the autoencoder are saved in feature store and will be used as input of the sequential model. |
| 8 | SEQNBT model is trained based on customer's transaction embeddings sequences from step 7. |
| 9 | Lookup service is used to query encoded transactions embeddings from the feature store. |
| 10 | SEQNBT model is deployed for real-time inference. |
| 11 | Identify Nearby Merchant is a microservice to filter nearby merchants based on geolocation and prediction of SEQNBT model. |
| 12 | Feature table containing all merchant geolocations is used by the service in step 11. |
| 13 | The prediction of customer's next best transaction will be delivered to the down-streaming services. |

## 5. CONCLUSIONS

In this paper, we proposed the SEQNBT model to predict both transaction amount and industry category of a user's next most likely transaction. Our method uses a stacked auto-encoder model to extract insightful transaction representations by incorporating more contextual information. When incorporated into the sequential model, the encodings and temporal features result in better performance compared to the pure GRU-based and Attention-based models. Our approach makes it possible to predict the amount of money a user is willing to pay for his or her next transaction. Further, the predicted SIC helps identify which industry the purchase will be in, so that we can identify a suitable merchant. We also presented a high-level architecture for delivering predictions to end users through EY Nexus product platform.

In the future, we anticipate driving SEQNBT through Nexus across a variety of data domains to measure business impact in a variety of use case contexts, not just within the offers space but also beyond into experience interaction management, as well.

156       Computer Science & Information Technology (CS & IT)## ACKNOWLEDGMENTS

The authors thank Kanagasabapathi Balasubramanian, Azeddine Kasmi, Jonathan DeGange, Kathleen M Calabro, Ryan D Battles, Lucy Bryans for project oversight and support. The views and conclusions expressed in this material are those of the authors and should not be interpreted as representing the official policies or endorsements, either expressed or implied, of Ernst & Young LLP.
## REFERENCES

[1] D. Jannach and M. Jugovac, "Measuring the business value of recommender systems," ACM Transactions on Management Information Systems (TMIS), vol. 10, no. 4, pp. 1-23, 2019.

[2] M. Sharma and S. Mann, "A survey of recommender systems: approaches and limitations," International Journal of Innovations in Engineering and Technology, vol. 2, no. 2, pp. 8-14, 2013.

[3] H. Fang, D. Zhang, Y. Shu, and G. Guo, "Deep learning for sequential recommendation: Algorithms, influential factors, and evaluations," ACM Transactions on Information Systems (TOIS), vol. 39, no. 1, pp. 1-42, 2020.

[4] D. Sculley et al., "Hidden technical debt in machine learning systems," Advances in neural information processing systems, vol. 28, 2015.

[5] C. C. Keith Strier, Jeanne Boillet, "How do you teach AI the value of trust?," 2018. [Online]. Available: https://assets.ey.com/content/dam/ey-sites/ey-com/en_gl/topics/digital/ey-how-do-you-teach-ai-the-value-of-trust.pdf

[6] F. Cacheda, V. Carneiro, D. Fernández, and V. Formoso, "Comparison of collaborative filtering algorithms: Limitations of current techniques and proposals for scalable, high-performance recommender systems," ACM Transactions on the Web (TWEB), vol. 5, no. 1, pp. 1-33, 2011.

[7] X. Amatriain and J. Basilico, "Recommender systems in industry: A netflix case study," in Recommender systems handbook: Springer, 2015, pp. 385-419.

[8] L. Morrison, "What's new in recommender systems," ed, 2020.

[9] S. Wang, L. Hu, Y. Wang, L. Cao, Q. Z. Sheng, and M. Orgun, "Sequential recommender systems: challenges, progress and prospects," arXiv preprint arXiv:2001.04830, 2019.

[10] L. Hu, J. Cao, G. Xu, L. Cao, Z. Gu, and C. Zhu, "Personalized recommendation via cross-domain triadic factorization," in Proceedings of the 22nd international conference on World Wide Web, 2013, pp. 595-606.

[11] J. Tang and K. Wang, "Personalized top-n sequential recommendation via convolutional sequence embedding," in Proceedings of the eleventh ACM international conference on web search and data mining, 2018, pp. 565-573.

[12] M. Ma et al., "Mixed Information Flow for Cross-domain Sequential Recommendations," ACM Transactions on Knowledge Discovery from Data (TKDD), vol. 16, no. 4, pp. 1-32, 2022.

[13] S. Zhang, Y. Tay, L. Yao, A. Sun, and J. An, "Next item recommendation with self-attentive metric learning," in Thirty-Third AAAI Conference on Artificial Intelligence, 2019, vol. 9.

[14] A. M. Elkahky, Y. Song, and X. He, "A multi-view deep learning approach for cross domain user modeling in recommendation systems," in Proceedings of the 24th international conference on world wide web, 2015, pp. 278-288.

[15] B. Hidasi, A. Karatzoglou, L. Baltrunas, and D. Tikk, "Session-based recommendations with recurrent neural networks," arXiv preprint arXiv:1511.06939, 2015.

[16] P. Covington, J. Adams, and E. Sargin, "Deep neural networks for youtube recommendations," in Proceedings of the 10th ACM conference on recommender systems, 2016, pp. 191-198.

[17] B. Hidasi, M. Quadrana, A. Karatzoglou, and D. Tikk, "Parallel recurrent neural network architectures for feature-rich session-based recommendations," in Proceedings of the 10th ACM conference on recommender systems, 2016, pp. 241-248.

[18] Q. Chen, H. Zhao, W. Li, P. Huang, and W. Ou, "Behavior sequence transformer for e-commerce recommendation in alibaba," in Proceedings of the 1st International Workshop on Deep Learning Practice for High-Dimensional Sparse Data, 2019, pp. 1-4.

[19] X. Xia et al., "TransAct: Transformer-based Realtime User Action Model for Recommendation at Pinterest," arXiv preprint arXiv:2306.00248, 2023.

## AUTHORS

**Xin Chen** is a Machine Learning Engineer of FSO Tech Consulting - Digital & Emerging Technology at Ernst & Young, LLP US, based out Boston, MA. office. She has been working on leveraging AI/ML skills to solve business problems in a variety of domain fields for 5 years post her Ph.D. graduation from Old Dominion University on Electrical and Computer Engineering. Xin got her MS degree in Optics from University of Shanghai for Science and Technology, China, and got her BS degree in Physics from Anhui Normal University, Anhui, China. Her interests in data science lie on the more technical side – specifically Machine Learning, Deep Learning, Natural Language Processing, Transformer models, and Recommendation system.

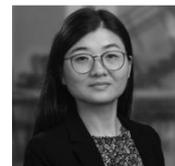

**Alex Reibman** is the former Machine Learning Engineer Manager of FSO Tech Consulting – Digital & Emerging Technology at Ernst & Young, LLP US, based out of the Hoboken, NJ office. He holds a degree in Economics and Philosophy from Emory University. He has a diverse background of experience and a focus on finding creative ways to solve problems with technology.

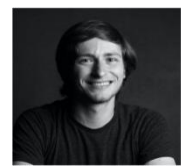

**Sanjay Arora** is CEO of Substep Technologies, Inc., a startup that helps people rethink their work life in the post Covid economy.  Sanjay holds undergraduate and master's degrees in information systems from Wayne State University and Carnegie Mellon University, respectively, and a PhD in science, technology, and innovation policy from the Georgia Institute of Technology. His past roles include AI/ML engineering leader and architect at EY and Senior Data Scientist for the American Institutes for Research.

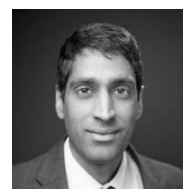